\documentclass[namedreferences]{SolarPhysics}
\usepackage[optionalrh]{spr-sola-addons} % For Solar Physics 
\usepackage{graphicx}        % For eps figures, newer & more powerfull
\usepackage{color}           % For color text: \color command
\usepackage{url}             % For breaking URLs easily trough lines
\begin{document}
\begin{article}
\begin{opening}
\title{The 10-rotation Periodicity in Sunspot Areas}
\author{R. \surname{Getko}}

\runningauthor{R. Getko}
\runningtitle{The 10-rotation Periodicity}

\institute{Astronomical Institute, University of Wroc{\l}aw, Poland,
email: \url{getko@astro.uni.wroc.pl}\\
}

\begin{abstract}
I study the sunspot area fluctuations over the epoch of 12 solar cycles (12-23). 
Lately, I found three significant quasi-periodicities at 10, 17 and 23 solar rotations, 
but two longer periods could be treated as subharmonics of the 10-rotation 
period. Thus, I search this period during the low- and the high-activity periods of each solar cycles. 
Because of the N-S asymmetry I consider each solar hemisphere separately. 
The skewness of each fluctuation probability distribution suggests that the positive and 
the negative fluctuations could be are examined separately. To avoid the problem when a few strong fluctuations could 
create an auto-correlation or a wavelet peak, I also analyse the transformations of fluctuations 
for which the amplitudes at the high- and the low-activity periods are almost the same. 
The auto-correlation and the wavelet analyses show that the 10-rotation period is mainly detected 
during the high-activity periods, but it also exists during a few low-activity periods.
\end{abstract}


\end{opening}

\section{Introduction}

In the last decades, the mid-term quasi-periodicities of many solar activity tracers have been discussed.
\citeauthor{wol} \shortcite{wol} reported the about $12$ rotations periodicity in the monthly Wolf number 
variations from 1749 to 1979. This periodicity was also found in many solar activity indicators. 
\citeauthor{del} \shortcite{del}  detected it in  the solar diameter measurements during cycle 21.\\ 
\citeauthor{lea} \shortcite{lea} found it in  the power spectrum  
of the sunspot blocking function, 10.7 cm radio flux, sunspot number and plage index daily data during the three 
solar cycles (Nos. 19-21). \citeauthor{pap} \shortcite{pap} showed that an 8-11 month period existed in the total 
and UV irradiances (1980 and 1982-1988 respectively), 10.7 radio flux (1947-1989), the Ca K plage index (1970-1987), 
the sunspot blocking function (1874-1982) and the Mg {\sc ii} core-wing ratio (1978-1986). \citeauthor{aki} \shortcite{aki} 
detected it in areas and numbers of sunspot groups from 1969 to 1986.  \citeauthor{oli} \shortcite{oli} studied daily 
sunspot numbers from 6 to 21 solar cycles. They adopted two different power spectrum methods and found a periodicity 
of about 323 days in three solar cycles only. They also found a significant periodicity at 350 days (13 rotations) 
during cycles 12-21 together and in some individual cycles.  \citeauthor{get} \shortcite{get} found 
statistically significant quasi-periodicity of about 9 months (10 rotations) in both high- and low-activity periods 
for the monthly Wolf number fluctuations from 1 to 22 solar cycles and 
for the group sunspot number fluctuations from 5 to 18 solar cycles.

Two longer periods at 17 rotations and at 23 rotations were found in many solar activity parameters.
More up to date review is by \citeauthor{ob} \shortcite{ob}. Lately, \citeauthor{get1} \shortcite{get1} 
showed that they could be treated as subharmonics of the 10-rotation period. This facts could explain a wide range 
of periodicities in various solar indices at all levels from the tachocline to the Earth.

In this paper, I search the 10-rotation period in the sunspot areas during 
the low- and the high-activity periods for cycles 12-23. 
The N-S asymmetry of solar activity (\citeauthor{viz}, \citeyear{viz}) suggests that both hemispheres should be considered 
separately. Because each empirical probability distribution of fluctuations is asymmetrical, 
the positive and the negative fluctuations are considered separately. To avoid 
the problem when a few strong fluctuations could create an auto-correlation or a wavelet peak, I transform 
each fluctuation time series into a new time series which has a constant variance (\citeauthor{get}, \citeyear{get}). 
The amplitudes of each new  time series at the high- and the low-activity periods are almost the same. 
The auto-correlation and the wavelet analyses of the original fluctuations and their transformations are used to find 
the mid-term periodicities from high- and low-activity periods in each solar cycle.

\section{The Stationary Version of Sunspot Area Fluctuations}

\begin{figure}
\centerline{\includegraphics[width=26pc]{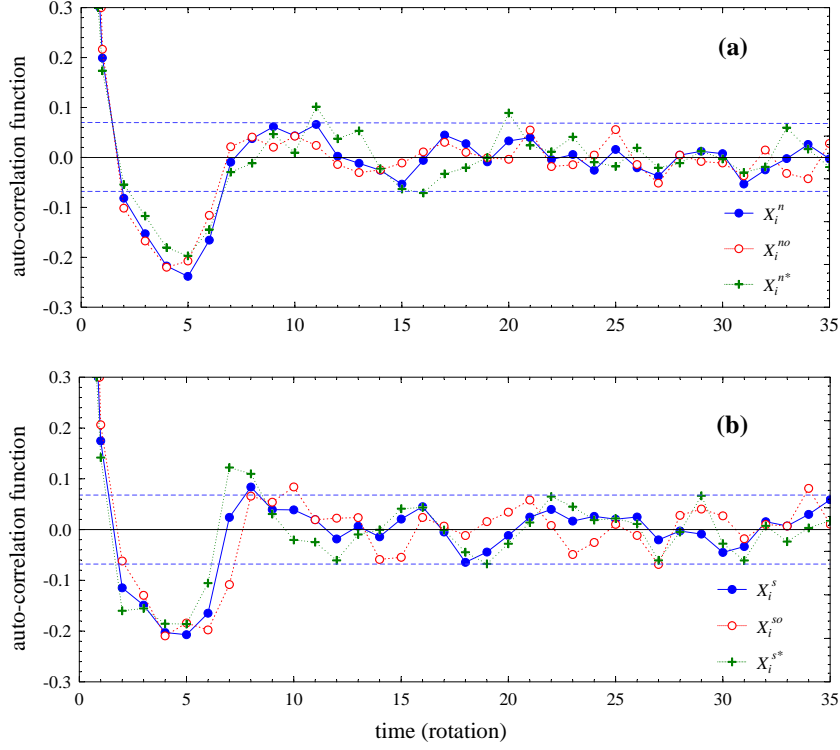}}
\caption[ ]{(a) Auto-correlation functions ($c_\tau$) of  $\{X_i^n\}$, $\{X_i^{no}\}$ and $\{X_i^{n*}\}$ (for all the data combined). 
The dashed horizontal lines represent two standard errors for each $c_\tau$ functions. 
(b) Same as for (a), but for  $\{X_i^s\}$, $\{X_i^{so}\}$ and $\{X_i^{s*}\}$.\\}  
\label{fig2}      
\end{figure}

I consider the daily sunspot areas for the northern hemisphere $(D_l^n)$, and the southern\quad hemisphere $(D_l^s)$ \quad 
for \quad  solar \quad  cycles 12-23\quad  available\quad at\quad 
the\quad \\National\quad Geophysical\quad Data\quad Center\quad (NGDC)\\(http://solarscience.msfc.nasa.gov/greenwch/). 
For the $i$-th Carrington rotation I evaluate the mean area for the northern hemisphere $(S_i^n)$:
$S_i^n=\frac{1}{K} \sum_{l=1}^{K}{D_l^n},$ where $K$ is the number of days for the $i$-th rotation. 
I define the fluctuation $(F_i^n)$ of the mean area $(S_i^n)$ from the smoothed mean area:
$F_i^n=S_i^n-\overline{S_i^n}\quad \mbox{for} \quad i=1,\ldots, N,$ where
$\overline{S_i^n}=\frac{1}{13} \sum_{j=i-6}^{i+6}{S_i^n}$.
Each of $\{F_i^n\}$ and $\{F_i^s\}$ contains $N=1706$ elements. 
The beginning and the end of each cycles is defined by the minima 
of the smoothed monthly Wolf numbers. Each time series has the zero mean value 
and the time-dependent variance. To stabilize the variance I apply the procedure which is shown by 
\citeauthor{get} \shortcite{get} in Appendix. For the $i$-th rotation the value $X_i^n$ of the new time series 
$\{X_i^n\}^{N}_{i=1}$  is a function of two parameters $k$ and $u$ (see equation (A2) in the Appendix). 
I estimate the values $k$ and $\log A$ from equation (A10). The best solution  is for $u=13$ (rotations) and 
$k=0.68\pm 0.02$. For $\{F_i^s\}$ it is for $u=13$ (rotations) and $k=0.83\pm 0.02$. Both the coefficients $k$ 
are statistically significant (the F-statistic is used). 

To verify the stationarity conditions of $\{X_i^n\}$ and $\{X_i^s\}$ the means 
($\overline{{\cal X}}={\frac{1}{u}\sum_{j=i_1-[\frac{u}{2}]}^{i_1+[\frac{u}{2}]} {X_j}}$), standard deviations\quad
(\boldmath $\overline{\sigma}$\unboldmath$=\sqrt{\frac{1}{u}\sum_{j=i_1-[\frac{u}{2}]}^{i_1+[\frac{u}{2}]} {({X_j}-\overline{{\cal X}})}^2}$ \unboldmath), 
and \\auto-correlation functions ($c_\tau$) are computed. 
For each time series  about $70\%$ of the values of $\overline{{\cal X}}$ for $u=13$ and $i_1=7, 20,\ldots, 1698$ 
belong to the interval $(\overline{{\cal X}}-{\rm \widehat{\sigma}}, \overline{{\cal X}}+\widehat{\sigma})$, 
where $\overline{{\cal X}^n}\approx -0.15$ and $\widehat{\sigma^n}\approx 0.7$, and 
$\overline{{\cal X}^s}\approx -0.08$, and $\widehat{\sigma^s}\approx 0.3$.
All sample standard deviations for each time series belong to the interval \boldmath $(\overline{\sigma}$\unboldmath$-2\widehat{\sigma}$, 
\boldmath$\overline{\sigma}$\unboldmath$+2\widehat{\sigma})$, where \boldmath $\overline{\sigma^n}$ \unboldmath $\approx 4.4$ and   
$\widehat{\sigma^n}\approx 0.4$ and  \boldmath $\overline{\sigma^s}$ \unboldmath $\approx 1.8$ and   
$\widehat{\sigma^s}\approx 0.2$. Figs.  \ref{fig2}a and \ref{fig2}b show the $c_\tau$ functions  
of  $\{X_i\}$, $\{X_i^o\}$ and $\{X_i^*\}$ for all 12 cycles for the northern and the southern hemispheres  
respectively. Here $\{X_i^o\}$ and $\{X_i^*\}$ are computed after dividing $\{X_i\}_{i=1}^{N}$ into two parts: 
those from the periods of high and low sunspot activities, respectively. To evaluate these periods the smoothed monthly Wolf numbers
$\{\overline{R_m}\}_{m=1}^{N_1}$ are used. All $m$ for which the values $\overline{R_m}$ are less (greater) than 
the mean value of $\{\overline{R_m}\}$ define low (high) activity periods. 
Two standard errors for each $c_{\tau}$ are plotted as the dashed horizontal lines.  Because each of three time series 
has a different length $M$, the bounds for the shortest time series are plotted ($\pm 0.068$).  
The differences between the $c_\tau$ functions of  $\{X_i^{no}\}$ and $\{X_i^{n*}\}$ are greater than $2\sigma$ for $\tau = 11, 13, 16, 17, 20, 25, 33, 34$
rotations. For  the southern hemisphere  both the $c_\tau$ functions significantly differ 
at $\tau =7, 10, 12, 15, 19, 23$, $30, 34$ rotations. Although the means and the standard deviations of $\{X_i\}$, $\{X_i^{o}\}$ and 
$\{X_i^*\}$ are aproximately constant the $c_\tau$ functions are different. Thus, $\{X_i^n\}$ and $\{X_i^s\}$ are not stationary.

Because each fluctuation probability distribution has positive skew (\citeauthor{get1}, \citeyear{get1}), 
the positive and the negative fluctuations are considered separately. 
For $\{F_i^n\}$ they can be defined as follows:
$$F_i^{n+}=\left\{\begin{array}{r@{\quad \mbox{where} \quad}l}
  0 & F_i^n\le 0 \\ F_i^n & F_i^n>0
  \end{array} \right. \quad \mbox{for} \quad i=1,\ldots,N.$$ 
$$F_i^{n-}=\left\{\begin{array}{r@{\quad \mbox{where} \quad}l}
  0 & F_i^n> 0 \\ F_i^n & F_i^n\le 0
  \end{array} \right. \quad \mbox{for} \quad i=1,\ldots,N. $$
The same definitions are applied to $X_i^{n+}$ and $X_i^{n-}$, and to all these time series for the southern hemisphere.
\begin{figure}
\centerline{\includegraphics[width=28pc]{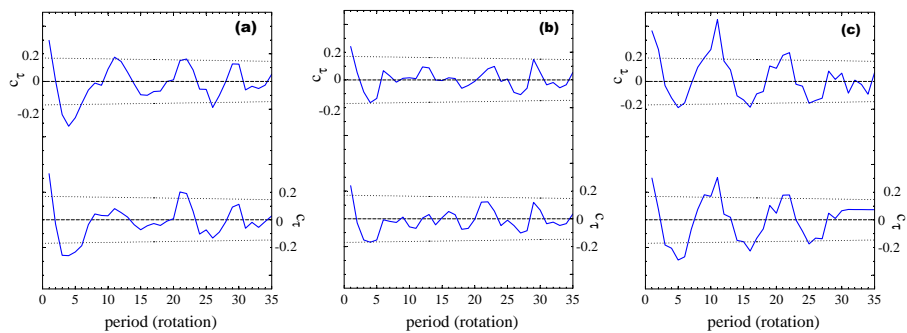}}
\caption[ ]{(a) \it{Top:} \rm Auto-correlation function ($c_\tau$) of $\{F_i^{n}\}$  for cycle 18. \it{Bottom} \rm Same as for the upper curve, but
for $\{X_i^{n}\}$. The dotted lines represent two standard errors of each $c_\tau$ function. 
   (b) \rm The same as for (a), but for $\{F_i^{n+}\}$ and $\{X_i^{n+}\}$. 
   (c) \rm The same as for (a), but for $\{F_i^{n-}\}$ and $\{X_i^{n-}\}$. 
\\}  
\label{fig4}      
\end{figure}

\section{The periodicities during high- and low-activity periods}

To find the periodicities  during high-activity periods I apply the auto-corre\-la\-tion function ($c_\tau$). 
The original, the positive and the negative fluctuations are considered separately. The same is for transformations. 
I divide each of them into 12 parts of length $L\approx 11$ years. For example, for the northern hemisphere I consider 
$\{F_i^n\}_{i=1}^{L}$, $\{X_i^n\}_{i=1}^{L}$, $\{F_i^{n+}\}_{i=1}^{L}$, $\{X_i^{n+}\}_{i=1}^{L}$, $\{F_i^{n-}\}_{i=1}^{L}$ and
$\{X_i^{n-}\}_{i=1}^{L}$ for each of 12 cycles. Figs. \ref{fig4}a-c show their  functions $c_\tau$ for cycle 18. 
For the $c_\tau$  of $\{F_i^n\}$ (Fig. \ref{fig4}a, top) a global maximum is greater than $2\sigma$ at $\tau=11$ rotations, 
for $\{X_i^n\}$ (bottom) the  value $c_{11}$ is less than $1\sigma$.   Moreover, for $\{F_i^{n+}\}$ the value $c_{12}$  is between 
$1\sigma$ and $2\sigma$ and for $\{X_i^{n+}\}$ it is less than $1\sigma$ (Fig. \ref{fig4}b). A decrease of $c_\tau$ at $\tau\in[7, 13]$ rotations, 
which is at least $1\sigma$, is in $71\%$ of 24 cases (12 cycles in each hemisphere). Because the transformation of $\{F_i^{n}\}$ 
into $\{X_i^{n}\}$ stabilizes the variance of $\{F_i^{n}\}$ such a decrease  is probably created by strong fluctuations from the high-activity period.
Moreover, the width of the $c_\tau$ maximum of $\{F_i^{n+}\}$ (evaluated for the $1\sigma$ level) is between two and five rotations. 
There are also a few cases (for example, for $\{F_i^{s+}\}$ cycles 13, 14, 18, 19, 20) for which each function $c_\tau$ 
has two or three peaks (their width for the $1\sigma$ level is one rotation). 
This means, that the strong fluctuations are quasi-periodical. Fig. \ref{fig4}c shows that for $\{F_i^{n-}\}$ and $\{X_i^{n-}\}$ in cycle 18 
both  $c_{11}$ are  greater than $2\sigma$. 
This could indicate that the fluctuations from the whole cycle create both peaks. In addition, a decrease of $c_{11}$ 
(it is at least $1\sigma$) could be created by fluctuations from the high-activity period. It is also found in $80\%$ of 24 cases.
The longer $c_\tau$ periods are treated as subharmonics of about 10-rotation period (\citeauthor{get1}, \citeyear{get1}).

\begin{figure}
\centerline{\includegraphics[width=28pc]{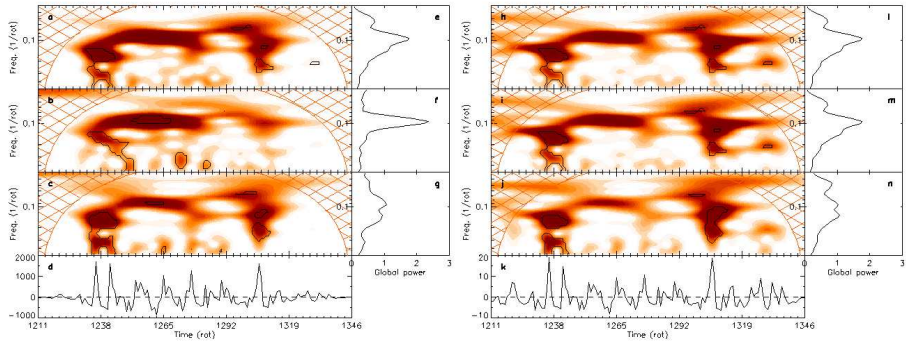}}
%\centerline{\includegraphics[width=26pc]{r3.ps}}
\caption{\bf {a-c:} \rm  Wavelet power spectra of \bf{a:} \rm $\{F_i^{n}\}$, \bf{b:} \rm$\{F_i^{n-}\}$ and \bf{c:} \rm $\{F_i^{n+}\}$ 
mapping a time-frequency evolution of about 10-rotation periodicity. 
Top values of wavelet power are denoted by gradual darkening. 
Black contours denote significance levels of 95 per cent for detected peaks. A cone of influence is marked by the dashed region. 
\bf{e-g:} \rm  Corresponding global wavelet power spectra.  \bf{d:} \rm Time series $\{F_i^{n}\}$ for cycle 18.
\bf{h-n:} \rm Same as \bf{a-g}, \rm but $\{X_i^{n}\}$, $\{X_i^{n-}\}$ and $\{X_i^{n+}\}$.}
\label{fig5}      
\end{figure}

I also apply the Morlet wavelet (\citeauthor{tor}, \citeyear{tor}) to six time series for each of 24 cases. 
Fig.  \ref{fig5}a (top) shows the local  spectrum of $\{F_i^{n}\}$ for cycle 18. 
Black solid contours denote the 95 per cent significance level for detected peaks. 
The global maximum of the integrated spectrum (right) at $\tau = 11$ rotations is mainly created by 
fluctuations from  the high-activity period. It confirms the auto-correlation result (Fig. \ref{fig2}a). 
Similar wavelet map is for $\{F_i^{n+}\}$ (bottom), but the peak at $\tau = 11$ rotations is significant. 
The global spectrum shows two almost the same peaks at $\tau = 11$ and $8$ rotations. 
The wavelet map of $\{X_i^{n}\}$ (Fig.  \ref{fig5}b, top) for the high-activity period is almost the same as for $\{F_i^{n}\}$. 
The global spectrum  for $\{X_i^{n+}\}$ is also almost the same as for $\{F_i^{n+}\}$, but  the wavelet power at $\tau = 11$ rotations 
is significant for $\{F_i^{n+}\}$ while for $\{X_i^{n+}\}$ it is not significant. 
For $\{F_i^{n-}\}$ and $\{X_i^{n-}\}$ in each wavelet map the strongest peak at $\tau=11$ rotations is significant (Figs.  \ref{fig5}a and \ref{fig5}b, middle). 
It  extends in time during the high-activity period and dominates the integrated spectrum. Moreover, the wavelet map of  $\{X_i^{n-}\}$ 
shows this periodicity during the declining portion of cycle 18 and 
smaller power at $\tau\approx 8$ rotations during the rising portion of cycle 18.
The same was done for all 24 cases. The wavelet and the auto-correlation results are similar for $\tau\in [7, 13]$ rotations (the correlation between 
them is about 0.9 for $87\%$, $72\%$, $92\%$, $61\%$ and $75\%$ of the $c_\tau$ peaks of $\{F_i^{n}\}$,  $\{F_i^{n+}\}$, 
 $\{F_i^{n-}\}$,  $\{X_i^{n}\}$  and $\{X_i^{n+}\}$ respectively).
For $\{X_i^{-}\}$ the correlation is about 0.8 for $83\%$ of $c_\tau$ peaks. 

To find the periodicities during the low-activity periods I examine the wavelets for 12 cycles together.
The map of $\{X_i^{n+}\}$ shows the significant peaks at about 10 rotations 
from these periods (cycles 13-14, 15-16, 17-18, 20-21). 
For $\{X_i^{n-}\}$ there are also several significant peaks (cycles 13-14, l6-17, 17-18, 18-19 and 20-21). 
Both maps of $\{X_i^{s+}\}$ and $\{X_i^{s-}\}$ also show significant peaks during these periods 
(cycles 12-13, 13-14, 15-16, 16-17, 18-19 and cycles 12-13, l5-16, 16-17 respectively). 

\section{Discussion}

The transformation of fluctuations into the new time series with a constant variance enables one to evaluate the time intervals where 
the functions $c_\tau$ have the most dominant periods. Such analyses indicate that the mean time distance between 
strong fluctuations from solar maxima is about 10 rotations for each hemisphere separately. Moreover, 
the division of $\{X_i\}$ for each hemisphere into two shorter parts:  containing data from 
low-activity periods ($\{X_i^*\}$) and  from high-activity periods ($\{X_i^o\}$) enables one to verify 
possible differences between periodicities existing during these periods in 12 solar cycles.

For each hemisphere the  functions $c_\tau$ of all three these time series 
have wide global maxima at $\tau\in[7, 13]$ rotations.
The functions $c_\tau$ of $\{X_i^n\}$ and $\{X_i^{n*}\}$ have the significant global maxima at $\tau=11$ rotations, but for $\{X_i^{no}\}$
two local maxima at $8$ and $10$ rotations  are between $1\sigma$ and $2\sigma$.
For the southern hemisphere all three global maxima are greater than $2\sigma$.
For $\{X_i^{so}\}$ this maximum contains two almost the same peaks at $\tau =8, 10$ rotations.
These results confirm the wavelets for all time series. All global spectra have the dominant peaks at $\tau=10$ rotations. 
For $\{X_i^{n+}\}$ and $\{X_i^{s+}\}$ these peaks are much wider and lower than for $\{X_i^{n-}\}$ and $\{X_i^{s-}\}$. 
These wide maxima indicate that the time separations between fluctuation maxima 
oscillate around the average length $\tau$. This fact could be caused by the differential rotation and 
the different rotation rate of activity complexes during their life time (\citeauthor{gau}, \citeyear{gau}).
In addition,  \citeauthor{get2} \shortcite{get2} shows that strong positive fluctuations are always created by 2-4 activity complexes which 
have not their largest size at the same time (the maxima of their contribution to the monthly Wolf number are shifted from one to three months).
Namely, the fluctuations are not strictly periodical, but quasi-periodical. 
Moreover, because the  functions $c_\tau$ (and the wavelets) 
of positive and of negative fluctuations are different, the positive fluctuations, 
which describe a strong magnetic flux emergence, are rather a sequence of pulses following at fixed time intervals 
than harmonic oscillations (\citeauthor{ob}, \citeyear{ob}). In addition, for 
$\{F_i^{n+}\}$ and $\{F_i^{s+}\}$ the significant wavelet peaks at 10 rotations usually exist 
during short time intervals of high-activity periods and also exist in the maps of $\{X_i^{n+}\}$ and $\{X_i^{s+}\}$ during several 
low-activity periods.

I also obtained similar results for the monthly Wolf numbers and for the group sunspot numbers during high-activity periods (\citeauthor{get}, \citeyear{get}). 
For periods around 11-12 rotations \citeauthor{pra} \shortcite{pra} showed a significant wavelet power of the daily Wolf numbers for short time intervals during several 
solar maxima. Taking into consideration that the solar activity fluctuations are quasi-periodical, the results obtained for daily data and data used 
in this paper are similar. Moreover, an analysis of the auto-correlation of daily sunspot areas for solar cycles 10-20 (\citeauthor{bog}, \citeyear{bog}) 
leads to the evidence for the existence of a few sites of unusually strong solar activity in each solar cycle that persist for about 10 rotations and 
generally rotate with a period of about 27 days.
\section{Conclusions}

The following results have been obtained: 
\begin{itemize}
\item [{\bf 1.}] The auto-correlation and the wavelet analyses of the original, the positive and the negative fluctuations and their transformations prefer 
the 10-rotation quasi-periodicity.
\item  [{\bf 2.}] The differences between the auto-correlation maxima (for $\tau\in [7, 13]$ rotations) of fluctuations and their transformations 
suggest that the 10-rotation period is created by strong fluctuations from high-activity periods. 
\item  [{\bf 3.}] The wavelet analysis confirms auto-correlation results for $\tau\in [7, 13]$ rotations (the correlation is about 0.9 for $87\%$, 
$72\%$, $92\%$, $61\%$ and $75\%$ of the auto-correlation peaks of $\{F_i^{n}\}$, $\{F_i^{n+}\}$, $\{F_i^{n-}\}$, $\{X_i^{n}\}$ and $\{X_i^{n+}\}$
respectively, and it is  0.8 for  $83\%$ of the auto-correlation peaks of $\{X_i^{n-}\}$). 
\item [{\bf 4.}] The wavelet maps of $\{X_i^{n+}\}$, $\{X_i^{n-}\}$, $\{X_i^{s+}\}$ and $\{X_i^{s-}\}$ for all 12 solar cycles together show several 
statistically significant peaks at about 10 rotations for low-activity periods. This means that during these periods small quasi-periodical fluctuations exist.
\end{itemize}

\end{article}
\end{document}